\documentclass[10pt,letterpaper]{article}
\usepackage{amsmath}
\usepackage{cite}
\usepackage{opex3}

\begin{document}
\title{Generating arbitrary photon-number entangled states for continuous-variable quantum informatics}
\author{Su-Yong Lee,$^{1}$ Jiyong Park,$^{1,2}$ Hai-Woong Lee,$^{2}$ and Hyunchul Nha$^{1,3,*}$}
\address{
$^{1}$Department of Physics, Texas A\&M University at Qatar, Education City, P.O.Box 23874, Doha, Qatar \\
$^{2}$Department of Physics, Korea Advanced Institute of Science and Technology, Daejeon 305-701, Korea \\
$^{3}$Institute f{\"u}r Quantenphysik, Universit{\"a}t Ulm, D-89069 Ulm, Germany
}
\email{$^{*}$hyunchul.nha@qatar.tamu.edu}

\begin{abstract}
We propose two experimental schemes that can produce an arbitrary photon-number entangled state (PNES) in a finite dimension. This class of entangled states naturally includes non-Gaussian continuous-variable (CV) states that may provide some practical advantages over the Gaussian counterparts (two-mode squeezed states).
We particularly compare the entanglement characteristics of the Gaussian and the non-Gaussian states in view of the degree of entanglement and the Einstein-Podolsky-Rosen correlation, and further discuss their applications to the CV teleportation and the nonlocality test.
The experimental imperfection due to the on-off photodetectors with nonideal efficiency is also considered in our analysis to show the feasibility of our schemes within existing technologies.
\end{abstract}


\section{Introduction}
Ever since Einstein-Podolsky-Rosen (EPR)'s argument against quantum mechanics was put forward \cite{1}, quantum entanglement has been a topic of great interest from a fundamental point of view. It can play a crucial role in manifesting striking differences between quantum and classical (e.g. local hidden-variable \cite{2, 3, 3-1}) descriptions of nature. Furthermore, it has also drawn much attention of practical interest because quantum correlations can be employed to carry out information tasks to the extent far beyond their classical counterparts, e.g. quantum computing \cite{4} and teleportation \cite{5}. In a bipartite setting, the primitive entangled states for discrete variables are the so-called Bell states, the maximally entangled states of two qubits, e.g. singlet state. In the regime of continuous variables (CVs), the Bell state can be realized in the form of two-mode squeezed state (TMSS), which becomes maximally entangled in the limit of infinite squeezing. The TMSS has been mostly the target entangled resource to produce for various quantum information tasks \cite{6} like CV quantum teleportation \cite{7}.  

The TMSS belongs to the class of Gaussian states, which has been extensively studied both theoretically and experimentally for CV quantum informatics \cite{8}. On the other hand, a great deal of attention has also been directed to the non-Gaussian regime (e.g. state engineering \cite{8-1} and characterization \cite{8-2,8-3}), as the non-Gaussian entangled states can provide some practical merits \cite{9-1, 9-2, 9-3, 9-4, 9-5, 9-6, 9-7, 9-8, 10} and even become an essential ingredient \cite{11-1, 11-2, 12-1, 12-2} for a number of quantum tasks.
Furthermore, when the quantum information processing is performed under realistic conditions, the quantum correlation is inevitably degraded and it thus becomes an important question whether Gaussian or non-Gaussian entangled states can be more robust against decoherence \cite{13,13-1}.
It was recently demonstrated that there exists a broad parameter space in which non-Gaussian entanglement can survive longer than Gaussian entanglement under noisy environments \cite{14-1, 14-2, 15} or quantum-limited amplifier \cite{16}.
With all these considered, it seems very desirable to have an experimental toolbox to generate a broad class of non-Gaussian entangled states in a controllable way.

In this paper we consider a class of CV entangled states in the photon-number entangled form $\sum^{N}_{n=0}C_n|n\rangle_a|n\rangle_b$, where $|n\rangle$ denotes a Fock state basis. One particular example is the TMSS with the coefficients $C_n=\lambda^n(1-\lambda^2)^{1/2}$ ($\lambda$: squeezing parameter, $N\rightarrow\infty$), which is the only Gaussian state among the photon-number entangled states. Another example is the pair-coherent state given by $C_n\sim\zeta^n/n!$ \cite{17, 18} which can be useful for a number of applications including quantum teleportation {\cite{19}, quantum metrology \cite{20}, and a Bell test \cite{21}.
In fact, a broad class of photon-number entangled states has been so far considered for the nonlocality test using homodyne detections \cite{22, 23, 24}.
Here we propose two experimental schemes to generate a finite-dimensional PNES with arbitrary coefficients,
$\sum^N_{n=0}C_n|n\rangle_a|n\rangle_b$ ($\sum^N_{n=0}|C_n|^2=1$), where the coefficients $C_n$ can be controlled with beam splitting and squeezing parameters. Both of our proposed schemes make use of coherent superposition operations in single-photon interferometic settings that erase the which-path information on the realized photonic operations.
The first scheme employs the second-order superposition operation $t\hat{a}\hat{a}^{\dag}+r\hat{a}^{\dag}\hat{a}$, which has been recently proposed and experimentally implemented in the context of proving bosonic commutation relation $[\hat{a},\hat{a}^{\dag}]=1$ \cite{25-1, 25-2, 26-1, 26-2}, together with two-mode squeezing operation. We note that the coherent operation $t\hat{a}\hat{a}^{\dag}+r\hat{a}^{\dag}\hat{a}$ was also discussed in the context of noiseless quantum amplifier \cite{26-3}.
On the other hand, the second scheme employs a sequence of nonlocal first-order coherent superpositions $t\hat{a}+r\hat{b}^{\dag}$. Its single-mode version $t\hat{a}+r\hat{a}^{\dag}$ was recently proposed for a quantum state engineering \cite{27} and also shown to be useful to enhance two-mode entanglement properties \cite{9-8, 29}.

We also address the usefulness of the finite-dimensional PNES for CV quantum teleportation \cite{7} and nonlocality test \cite{30-1, 30-2} compared with the two-mode squeezed state. Furthermore, it was very recently shown that the photon-number entangled states in finite dimension, e.g. $C_0|0\rangle_a|0\rangle_b+C_1|1\rangle_a|1\rangle_b$, can survive longer under noisy environments than the TMSS with the same degree of entanglement or energy \cite{14-1, 14-2}.
Therefore, our proposed schemes can be a useful tool not only for CV quantum applications but also for fundamental tests of quantum physics.

This paper is organized as follows. In Sec. 2, we first compare the entanglement properties of finite-dimensional PNES (non-Gaussian) and a two-mode squeezed state (Gaussian) in view of the degree of entanglement and the EPR correlation. In Sec. 3, we further investigate the usefulness of the PNES for CV quantum teleportation and nonlocality test. Then, we propose two experimental schemes to generate an arbitrary PNES, $\sum^N_{n=0}C_n|n\rangle_a|n\rangle_b$, in Sec. 4.  We illustrate the feasibility of our schemes in Sec. 5 by investigating the generation of a PNES up to two-photon correlation, i.e. $C_0|0\rangle_a|0\rangle_b+C_1|1\rangle_a|1\rangle_b+C_2|2\rangle_a|2\rangle_b$, considering realistic experimental conditions. 
In Sec. 6, our results are summarized.

\section{Entanglement and EPR correlation}\label{Entanglement and EPR correlation}
First, we briefly compare the TMSS and the PNES in terms of entanglement properties in order to identify the practical relevance of the PNES for CV quantum informatics.
For a pure two-mode state $|\Psi\rangle_{AB}$, the degree of entanglement can be quantified by the von Neumann entropy $E(\rho_A)=-{\rm Tr}_A[\rho_A\log_2{\rho_A}]$ for the reduced density operator $\rho_A={\rm Tr}_B[|\Psi\rangle_{AB}\langle\Psi|_{AB}]$.
For the class of photon-number entangled states $\sum^N_{n=0}C_n|n\rangle_a|n\rangle_b$, the von Neumann entropy becomes maximal when all the coefficients $C_n$ are identical. Thus, for the case of TMSS with $C_n=\lambda^n(1-\lambda^2)^{1/2}$, the state can have an infinite degree of entanglement with infinite squeezing, i.e. $\lambda=1$, which is practically impossible to achieve.
On the other hand, the finite-dimensional PNES can match or even surpass a finitely-squeezed TMSS as an entangled resource.
In Fig. 1(a), we plot the degree of entanglement for the TMSS (blue solid) as a function of the squeezing parameter $s=\tanh^{-1}\lambda$.
This is compared with the maximal possible entanglement for the PNES with equal coefficients ($C_1=C_2=\dots=C_N$) of dimensions $N=1$ (red dotted), 2 (red dashed), and 10 (red dot-dashed). The degrees of entanglement for the PNESs are given by $1,~1.585$ and $3.459$, respectively. To achieve such degrees of entanglement, the squeezing of the TMSS should be $s=0.5185~(4.506~dB),~s=0.7335~(6.374~dB)$ and $s=1.391~(12.09~dB)$, respectively. In the pulsed-regime generation of squeezed states, the level of squeezing currently available from an optical parametric amplifier is $s=0.403~(3.5~dB)$ \cite{31,31-1} so that the PNES with $N=1$ can already surpass the entanglement of the TMSS.

We also look into another entanglement property, the EPR correlation, which is the total variance of a pair of EPR-like operators, ${\rm EPR}\equiv\Delta^2(\hat{x}_A-\hat{x}_B)+\Delta^2(\hat{p}_A+\hat{p}_B)$. Here $\hat{x}_j=\frac{1}{\sqrt{2}}(\hat{a}_j+\hat{a}^{\dag}_j)$ and $\hat{p}_j=\frac{1}{i\sqrt{2}}(\hat{a}_j-\hat{a}^{\dag}_j)$ $(j=A,B)$ are the quadrature amplitudes of the field that can be measured in homodyne detection.
The value of ${\rm EPR}$ below 2 represents the quantum correlation between the quadrature amplitudes of two modes.
In Fig. 1(b), the EPR correlations of the PNESs for the dimensions $N=1,~2,$ and $10$ are $1.172,~0.8315,$ and $0.2516$, respectively.
The corresponding levels of squeezing for the TMSS are given by $s=0.2674~(2.324~dB),~s=0.4388~(3.813~dB)$ and $s=1.037~(9.008~dB)$.
Thus the PNES with $N\geq 2$ can surpass the currently available TMSS ($s=0.403$) in view of the EPR correlation.

\begin{figure}
\centering\includegraphics[width=\textwidth]{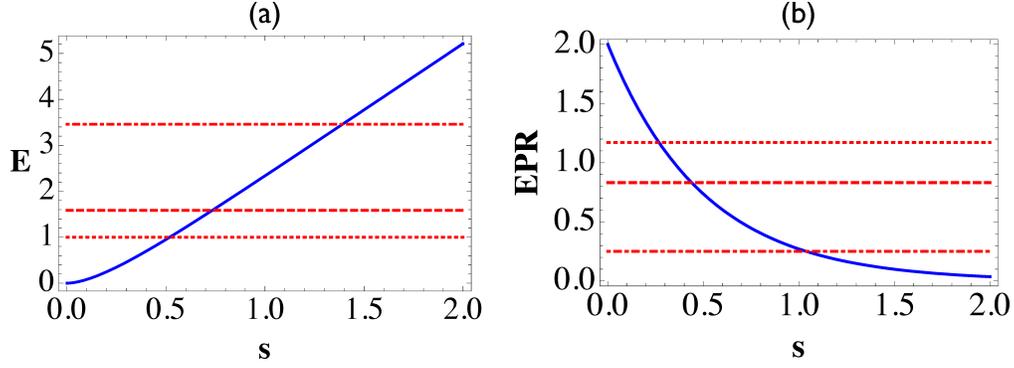}
\caption{(a) Degree of entanglement and (b) EPR correlation for the states: $|TMSS\rangle$ (blue solid) as a function of the squeezing parameter $s$, and $\sum^N_{n=0}C_n|n\rangle_a|n\rangle_b$ at $N=1$ (red dotted), $N=2$ (red dashed), $N=10$ (red dot-dashed). }
\label{FIG_1}
\end{figure}


\section{Applications: CV teleportation and nonlocality test}\label{Application: CV teleportation, and nonlocality test}
In this section, we further investigate the usefulness of a finite-dimensional PNES particularly for continuous variable (CV) teleportation and nonlocality test. 
For this purpose, we evaluate the quality of CV teleportation by the average fidelity between an unknown input state and the teleported state \cite{7},
and investigate the nonlocality test by Banaszek and W\'odkiewicz based on the phase-space distribution functions \cite{30-1, 30-2}.

{\bf (i)} The teleportation fidelity in the Braunstein-Kimble (BK) scheme \cite{7} can be evaluated in terms of the characteristic functions of an input state and its teleported state as
\begin{equation}
F=\frac{1}{\pi}\int d^2\lambda C_{\rm out}(\lambda)C_{\rm in}(-\lambda),
\end{equation}
where $C_{\rm out}(\lambda)=C_{\rm in}(\lambda)C_E(\lambda^*, \lambda)$ \cite{32}. Here $C_E(\lambda^*, \lambda)$ is the characteristic function of a two-mode entangled state. We consider the finite-dimensional PNES $\sum^N_{n=0}C_n|n\rangle_a|n\rangle_b$ for the dimensions $N=1$,$2$ and $3$ as an entangled resource.
For instance, the characteristic function of the PNES for $N=2$ is given by
\begin{align}
C_{E}(\lambda_2, \lambda_3)
=&e^{-(|\lambda_2|^2+|\lambda_3|^2)/2}[|C_0|^2+|C_1|^2(1-|\lambda_2|^2)(1-|\lambda_3|^2)\nonumber\\
&+\frac{|C_2|^2}{4}(2-4|\lambda_2|^2+|\lambda_2|^4)(2-4|\lambda_3|^2+|\lambda_3|^4)\nonumber\\
&+C^*_0C_1\lambda^*_2\lambda^*_3+C_0C^*_1\lambda_2\lambda_3+\frac{C^*_0C_2}{2}\lambda^{*2}_2\lambda^{*2}_3
+\frac{C_0C^*_2}{2}\lambda^{2}_2\lambda^{2}_3\nonumber\\
&+\frac{1}{2}(C^*_1C_2\lambda^*_2\lambda^*_3+C_1C^*_2\lambda_2\lambda_3)(|\lambda_2|^2-2)(|\lambda_3|^2-2)],
\end{align}
where $|C_0|^2+|C_1|^2+|C_2|^2=1$. For the case of teleporting an arbitrary coherent-state input, we find,
by optimizing the fidelity (1) using Eq. (2),
that the average fidelity can be achieved up to $F=0.7334$ at the choice of $C_0\approx 0.765$, $C_1\approx 0.535$ and $C_2\approx 0.359$. 

In Fig. 2(a), we compare the teleportation fidelity achieved via the PNES and the fidelity via the TMSS.
The optimal fidelity via each PNES at $N=1$, $2$ and $3$ corresponds to the fidelity via the TMSS with the squeezing
parameters $s=0.320$ ($2.776~dB$), $s=0.506$ ($4.397~dB$), and $s=0.638$ ($5.548~dB$), respectively.
Thus, the PNES at $N=2$ can surpass the fidelity via the TMSS with the currently available squeezing in the pulsed regime, i.e.,
$s\approx 0.403~(3.5~dB)$ \cite{32}.
As we show in Section 4, our proposed schemes do not require a high-level of squeezing to produce the optimal PNES (case of $N=2$) for CV teleportation.

\begin{figure}
\centering\includegraphics[width=\textwidth]{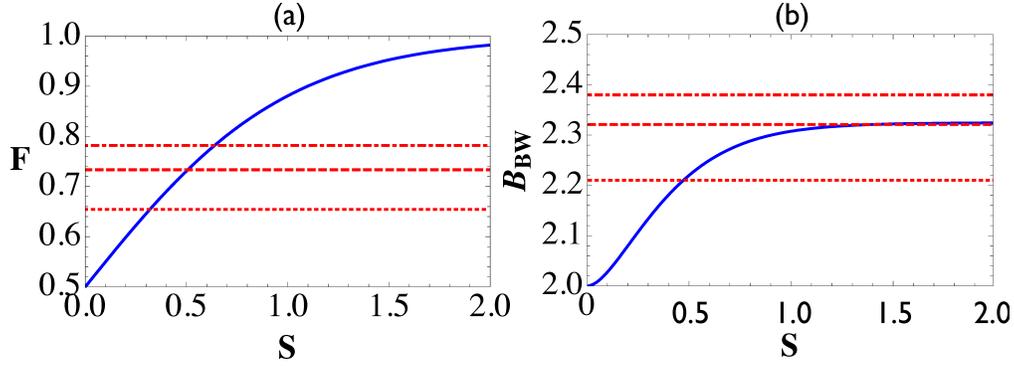}
\caption{(a) Average fidelity in teleporting a coherent state and (b) Bell parameter $B_{\rm BW}$ as a function of the squeezing parameter $s$ for the   $|TMSS\rangle$ (blue solid) and the PNES $\sum^N_{n=0}C_n|n\rangle_a|n\rangle_b$ at $N=1$ (red dotted), $N=2$ (red dashed)
and $N=3$ (red dot-dashed). The coefficients of the PNESs are optimized for each $N$. }
\label{FIG_5}
\end{figure}

{\bf (ii)} We next consider the nonlocality test by Banaszek and W\'odkiewicz that is addressed in phase space using the two-mode Wigner function \cite{30-1, 30-2}.
This Bell inequality is given by
\begin{equation}
B_{\rm BW}=\frac{\pi^2}{4}|W(\alpha,\beta)+W(\alpha,\beta^{'})+W(\alpha^{'},\beta)-W(\alpha^{'},\beta^{'})| \leq  2,
\end{equation}
where $W(\alpha,\beta)$ is the two-mode Wigner function.
We find that for the PNES, $\sum^N_{n=0}C_n|n\rangle_a|n\rangle_b$ at $N=2$, the Bell inequality can be violated up to $B_{\rm BW}=2.32088$ with the coefficients $C_0\approx 0.589$, $C_1\approx 0.700$ and $C_2\approx 0.404$. In Fig. 2(b), we see that this degree of nonlocality almost reaches the level for the TMSS with infinite squeezing \cite{33}.
Furthermore, we also see that the Bell violation by the PNES at $N=3$ surpasses the value $B_{\rm BW}$ of the TMSS in the entire region of squeezing.
We can again achieve such degree of Bell violation using the weak squeezing regime in our schemes (Sec. 4).

\section{Experimental schemes}\label{Implementation}
We now propose two optical schemes to generate an arbitrary PNES, $\sum^N_{n=0}C_n|n\rangle_a|n\rangle_b$. One scheme is based on a second-order coherent superposition operation with two-mode squeezing operations, and the other on a sequence of first-order coherent superposition operations.

{\bf (i)} We first consider the operation $t\hat{a}\hat{a}^{\dag}+r\hat{a}^{\dag}\hat{a}$ acting on a single-mode $a$, which is the coherent superposition of two product operations---photon addition followed by subtraction $(\hat{a}\hat{a}^{\dag})$ and photon subtraction followed by addition $(\hat{a}^{\dag}\hat{a})$. This coherent operation was experimentally implemented to prove the bosonic commutation relation $[\hat{a},\hat{a}^{\dag}]=1$ \cite{25-1, 25-2}.
While such a commutator is addressed as an equal superposition of the two product operations, $\hat{a}\hat{a}^{\dag}-\hat{a}^{\dag}\hat{a}$, we adopt an arbitrarily weighted superposition of the two operations, i.e., $t\hat{a}\hat{a}^{\dag}+r\hat{a}^{\dag}\hat{a}$.
In particular, we show that the single-mode operation $t\hat{a}\hat{a}^{\dag}+r\hat{a}^{\dag}\hat{a}$ together with two-mode squeezing operations $\hat{S}_{ab}(\xi)=\exp(-\xi\hat{a}^{\dag}\hat{b}^{\dag}+\xi^*\hat{b}\hat{a})$ can constitute an essential building block to generate an arbitrary PNES.

Suppose that a two-mode squeezing $\hat{S}_{ab}(\xi)$, the coherent operation $t\hat{a}\hat{a}^{\dag}+r\hat{a}^{\dag}\hat{a}$ and the inverse squeezing $\hat{S}^{\dag}_{ab}(\xi)$ are sequentially applied to an input state. That is, we apply a sequence of operations defined by
\begin{align}
\hat{O}_n\equiv&\hat{S}^{\dag}_{ab}(\xi_n)(t_n\hat{a}\hat{a}^{\dag}+r_n\hat{a}^{\dag}\hat{a})\hat{S}_{ab}(\xi_n)\nonumber\\
=&A_n+(t_n+r_n)(\hat{a}^{\dag}\hat{a}\cosh^2{s_n}+\hat{b}^{\dag}\hat{b}\sinh^2{s_n})\nonumber\\
&-(t_n+r_n)\cosh{s_n}\sinh{s_n}[\exp(-i\varphi_n)\hat{a}\hat{b}+\exp(i\varphi_n)\hat{a}^{\dag}\hat{b}^{\dag}],
\end{align}
where
\begin{equation}
A_n=t_n\cosh^2{s_n}+r_n\sinh^2{s_n},
\end{equation}
with $|t_n|^2+|r_n|^2=1$. In the above Eq. (4), the identity $\hat{S}^{\dag}_{ab}(\xi)\hat{a}\hat{S}_{ab}(\xi)=\hat{a}\cosh{s}-\hat{b}^{\dag}\exp(i\varphi)\sinh{s}$
and $\hat{S}^{\dag}_{ab}(\xi)\hat{a}^{\dag}\hat{S}_{ab}(\xi)=\hat{a}^{\dag}\cosh{s}-\hat{b}\exp(-i\varphi)\sinh{s}$ are used, where $\xi\equiv s\exp(i\varphi)$ \cite{34}.

When a vacuum state $|0\rangle_a|0\rangle_b$ is injected as an input, $\hat{O}_1$ yields a superposition of number states as $\hat{O}_1|0\rangle_a|0\rangle_b=\cosh^2{s_1}[(t_1+r_1\tanh^2{s_1})|0\rangle_a|0\rangle_b-\exp(i\varphi_1)(t_1+r_1)\tanh{s_1}|1\rangle_a|1\rangle_b]$. In principle, a succession of $\hat{O}_n$ applied on the vacuum states, $\prod^N_{n=1}\hat{O}_n|0\rangle_a|0\rangle_b$, can yield an arbitrary superposition state $\sum^N_{n=0}C_n|n\rangle_a|n\rangle_b$ by properly choosing the parameters $s_n$, $r_n$, $t_n$ and $\varphi_n$.
For instance, the state $\hat{O}_1|0\rangle_a|0\rangle_b\sim C_0|0\rangle_a|0\rangle_b+C_1|1\rangle_a|1\rangle_b$ can have a larger proportion of  $|1\rangle_a|1\rangle_b$, i.e. $|C_0|<|C_1|$,  under the condition $r_1\tanh{s_1}> t_1$.
For comparison, if one instead applies the original quantum scissor scheme on the TMSS that projects an input onto the subspace spanned by $|0\rangle$ and $|1\rangle$ \cite{35}, the output state becomes $\sim|0\rangle_a|0\rangle_b+\lambda|1\rangle_a|1\rangle_b$. That is, the vacuum state $|0\rangle_a|0\rangle_b$ is always more weighted than the single-photon state $|1\rangle_a|1\rangle_b$. 
On the other hand, a generalized scissor scheme proposed for the noiseless quantum amplifier \cite{35-1} can be used to control each coefficient arbitrarily in the $|0\rangle$-$|1\rangle$ subspace, which will be briefly discussed in Sec. 5.


\begin{figure}
\centering\includegraphics[scale=0.8]{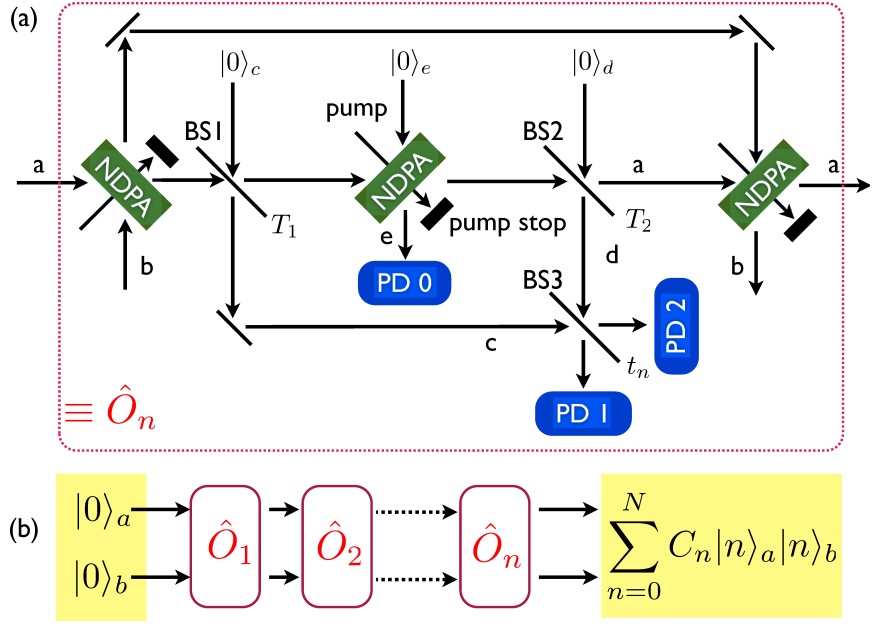}
\caption{(a) Experimental scheme to implement the operation $\hat{S}^{\dag}_{ab}(\xi)(t\hat{a}\hat{a}^{\dag}+r\hat{a}^{\dag}\hat{a})\hat{S}_{ab}(\xi)$ on an arbitrary state. BS1, BS2, and BS3 are beam splitters with transmissivities $T_1$, $T_2$ and $t_n$, respectively. PD0, PD1 and PD2: photo detectors. The operation is successfully achieved under the detection of a single photon at only one of two detectors PD1 and PD2, with PD0 clicked. (b) For a vacuum input state, the sequence of operations $\hat{O}_n$ can yield a finite dimensional PNES, $\sum^N_{n=0}C_n|n\rangle_a|n\rangle_b$. }
\label{FIG_3}
\end{figure}

The elementary operation $\hat{O}_n$ can be experimentally realized as depicted in Fig. 3(a).
An input state $|\psi\rangle_{ab}$ is first injected into a nondegenerate parametric amplifier (NDPA) with coupling parameter $\xi_n$, and then into the beam splitter BS1 (transmittance: $T_1\approx 1$) with the other input mode $c$ in a vacuum. This can be described by
\begin{equation}
\hat{B}_{ac}\hat{S}_{ab}(\xi_n)|\psi\rangle_{ab}|0\rangle_c
\approx (1-\frac{R^*_1}{T_1}\hat{a}\hat{c}^{\dag})\hat{S}_{ab}(\xi_n)|\psi\rangle_{ab}|0\rangle_c.
\end{equation}
Next the mode $a$ is further injected into another NDPA with small coupling $s\ll 1$ and the output is kept only under the condition of single-photon detection at PD0. That is,
\begin{equation}
_e\langle 1| \hat{S}_{ae}(1-\frac{R^*_1}{T_1}\hat{a}\hat{c}^{\dag})\hat{S}_{ab}(\xi_n)|\psi\rangle_{ab}|0\rangle_c|0\rangle_e \approx -s\hat{a}^{\dag}(1-\frac{R^*_1}{T_1}\hat{a}\hat{c}^{\dag})\hat{S}_{ab}(\xi_n)|\psi\rangle_{ab}|0\rangle_c,
\end{equation}
which is then injected into the BS2 ($T_2\approx 1$),
\begin{align}
&(-s)\hat{B}_{ad}\hat{a}^{\dag}(1-\frac{R^*_1}{T_1}\hat{a}\hat{c}^{\dag})\hat{S}_{ab}(\xi_n)|\psi\rangle_{ab}|0\rangle_{cd} \nonumber\\
&\approx (-s)(1-\frac{R^*_2}{T_2}\hat{a}\hat{d}^{\dag})\hat{a}^{\dag}(1-\frac{R^*_1}{T_1}\hat{a}\hat{c}^{\dag})\hat{S}_{ab}(\xi_n)|\psi\rangle_{ab}|0\rangle_{cd},
\end{align}
where $|0\rangle_{cd}=|0\rangle_c|0\rangle_d$.
The next beam splitter BS3 making the transformations $\hat{c}\rightarrow t_n\hat{c}+r_n\hat{d}$ and
$\hat{d}\rightarrow t_n\hat{d}-r_n\hat{c}$ gives
\begin{equation}
|S_{|\psi\rangle}\rangle\equiv(-s)[1-\frac{R^*_2}{T_2}\hat{a}(t_n\hat{d}^{\dag}-r_n\hat{c}^{\dag})]\hat{a}^{\dag}[1-\frac{R^*_1}{T_1}\hat{a}(t_n\hat{c}^{\dag}+r_n\hat{d}^{\dag})]\hat{S}_{ab}(\xi_n)|\psi\rangle_{ab}|0\rangle_{cd}.
\end{equation}
On detecting a single photon at PD1 (PD2) and no photon at PD2 (PD1), the state is projected to
$(t\hat{a}\hat{a}^{\dag}+r\hat{a}^{\dag}\hat{a})\hat{S}_{ab}(\xi_n)|\psi\rangle_{ab}$, where $t\sim \frac{R^*_2}{T_2}st_n$
and $r\sim \frac{R^*_1}{T_1}sr_n$ ($t\sim-\frac{R^*_2}{T_2}sr_n$ and $r\sim \frac{R^*_1}{T_1}st_n$). Finally, the NDPA with the coupling parameter $-\xi_n$ yields $|\psi\rangle_{\rm out}\sim \hat{S}_{ab}^{\dag}(\xi_n)(t\hat{a}\hat{a}^{\dag}+r\hat{a}^{\dag}\hat{a})\hat{S}_{ab}(\xi_n)|\psi\rangle_{ab}$, with the identity $\hat{S}_{ab}^{\dag}(\xi_n)=\hat{S}_{ab}(-\xi_n)$.

{\bf (ii)} Second, we show that the sequence of two first-order coherent superposition operations,
$({\rm t_{2n}}\hat{a}+{\rm r_{2n}}\hat{b}^{\dag})({\rm t_{2n-1}}\hat{b}+{\rm r_{2n-1}}\hat{a}^{\dag})$,
can also yield an operation similar to $\hat{O}_n$ in Eq. (4). A similar type of coherent operation was previously investigated in a form acting on a single-mode, $t\hat{a}+r\hat{a}^{\dag}$, which is the superposition of photon subtraction and addition \cite{27}.
Here we consider a {\it nonlocal} coherent superposition acting on two modes, $t\hat{a}+r\hat{b}^{\dag}$ ($t\hat{b}+r\hat{a}^{\dag}$).

We define an operator
\begin{align}
\hat{O}^{'}_n \equiv& ({\rm t_{2n}}\hat{a}+{\rm r_{2n}}\hat{b}^{\dag})({\rm t_{2n-1}}\hat{b}+{\rm r_{2n-1}}\hat{a}^{\dag})\nonumber\\
=& {\rm t_{2n-1}t_{2n}}\hat{a}\hat{b}+{\rm r_{2n-1}r_{2n}}\hat{a}^{\dag}\hat{b}^{\dag}
+{\rm r_{2n-1}t_{2n}}\hat{a}\hat{a}^{\dag}+{\rm t_{2n-1}r_{2n}}\hat{b}^{\dag}\hat{b},
\end{align}
where $|{\rm t_{2n-1}}|^2+|{\rm r_{2n-1}}|^2=1$ and $|{\rm t_{2n}}|^2+|{\rm r_{2n}}|^2=1$.
Given a vacuum state as an input, $\hat{O}^{'}_1$ yields a superposition of number states as
$\hat{O}^{'}_1|0\rangle_a|0\rangle_b= {\rm r_1}({\rm t_2}|0\rangle_a|0\rangle_b+{\rm r_2}|1\rangle_a|1\rangle_b)$.
Furthermore, a succession of $\hat{O}^{'}_n$, i.e., $\prod^N_{n=1}\hat{O}^{'}_n|0\rangle_a|0\rangle_b$, can yield any desired superposition state $\sum^N_{n=0}C_n|n\rangle_a|n\rangle_b$ by properly choosing the parameters ${\rm r_{2n-1}}$, ${\rm t_{2n-1}}$, ${\rm r_{2n}}$ and ${\rm t_{2n}}$.
Here the coefficients can be readily controlled only by the beam-splitter parameters as shown below. 

The operation $\hat{O}^{'}_n$ can be implemented as depicted in Fig. 4.
First, an arbitrary two-mode state $|\psi\rangle_{ab}$ is injected into an NDPA with small coupling
${\rm s_1}\ll 1$ and a BS1 with high transmissivity ${\rm T_1}\approx 1$, with mode $a$ ($b$) into NDPA (BS1). The other input modes to the NDPAs and the BSs are all in vacuum states.
Then, the BS3 (transmissivity: ${\rm t_{2n-1}}$) yielding the transformations $\hat{c}^{\dag}\rightarrow {\rm t_{2n-1}}\hat{c}^{\dag}+{\rm r_{2n-1}}\hat{d}^{\dag}$ and
$\hat{d}^{\dag}\rightarrow {\rm t_{2n-1}}\hat{d}^{\dag}-{\rm r_{2n-1}}\hat{c}^{\dag}$ gives the output
\begin{equation}
[1-\frac{\rm R^*_1}{\rm T_1}\hat{b}({\rm t_{2n-1}}\hat{d}^{\dag}-{\rm r_{2n-1}}\hat{c}^{\dag})][1-{\rm s_1}\hat{a}^{\dag}({\rm t_{2n-1}}\hat{c}^{\dag}+{\rm r_{2n-1}}\hat{d}^{\dag})]
|\psi\rangle_{ab}|0\rangle_{cd}.
\end{equation}
With the detection of single photon at PD1 (PD2) and no photon at PD2 (PD1), we see from Eq. (11) that the state is projected  to
$|\Phi\rangle_{ab}\equiv({\rm t^{'}_{2n-1}}\hat{b}+{\rm r^{'}_{2n-1}}\hat{a}^{\dag})|\psi\rangle_{ab}$, where ${\rm t^{'}_{2n-1}}\sim -\frac{\rm R^*_1}{\rm T_1}{\rm t_{2n-1}}$ ($\frac{\rm R^*_1}{\rm T_1}{\rm r_{2n-1}}$) and
${\rm r^{'}_{2n-1}}\sim -{\rm s_1r_{2n-1}}$ ($\rm -s_1t_{2n-1}$). Next, the output state is further injected into another NDPA with small coupling
${\rm s_2}\ll 1$ and a BS2 with high transmissivity ${\rm T_2}\approx 1$, with mode a (b) into BS2 (NDPA).
Finally, a beam splitter BS4 (transmissivity: ${\rm t_{2n}}$) yielding the transformations $\hat{e}^{\dag}\rightarrow {\rm t_{2n}}\hat{e}^{\dag}+{\rm r_{2n}}\hat{f}^{\dag}$ and $\hat{f}^{\dag}\rightarrow {\rm t_{2n}}\hat{f}^{\dag}-{\rm r_{2n}}\hat{e}^{\dag}$ gives
\begin{equation}
|S^{'}_{|\psi\rangle}\rangle \equiv[1-\frac{\rm R^*_2}{\rm T_2}\hat{a}({\rm t_{2n}}\hat{e}^{\dag}+{\rm r_{2n}}\hat{f}^{\dag})]
[1-{\rm s_2}\hat{b}^{\dag}({\rm t_{2n}}\hat{f}^{\dag}-{\rm r_{2n}}\hat{e}^{\dag})]|\Phi\rangle_{ab}|0\rangle_{ef}.
\end{equation}
Once again, with the detection of single photon at PD3 (PD4) and no photon at PD4 (PD3), we see from Eq. (12) that the state is projected  to
$({\rm t^{'}_{2n}}\hat{a}+{\rm r^{'}_{2n}}\hat{b}^{\dag})({\rm t^{'}_{2n-1}}\hat{b}+{\rm r^{'}_{2n-1}}\hat{a}^{\dag})|\psi\rangle_{ab}$, where
${\rm t^{'}_{2n}}\sim -\frac{\rm R^*_2}{\rm T_2}{\rm r_{2n}}$ ($-\frac{\rm R^*_2}{\rm T_2}{\rm t_{2n}}$) and ${\rm r^{'}_{2n}}\sim -{\rm s_2t_{2n}}$ (${\rm s_2r_{2n}}$).

\begin{figure}
\centering\includegraphics[scale=0.6]{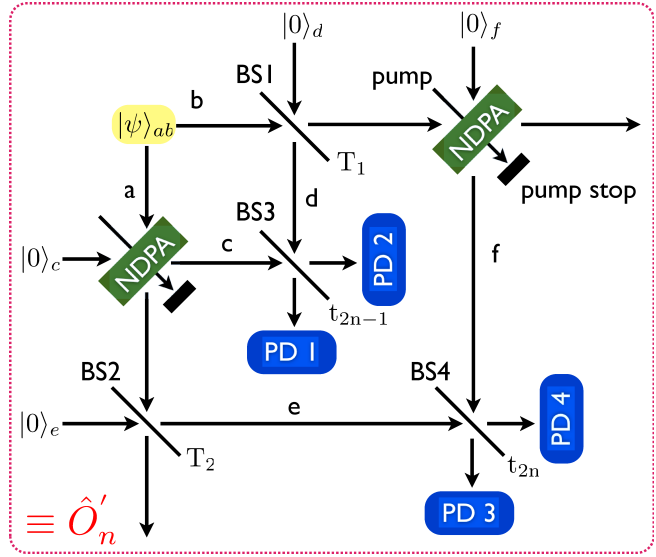}
\caption{Experimental scheme to implement the operation $({\rm t_{2n}}\hat{a}+{\rm r_{2n}}\hat{b}^{\dag})({\rm t_{2n-1}}\hat{b}+{\rm r_{2n-1}}\hat{a}^{\dag})$ on an input state $|\psi\rangle_{ab}$. BS1, BS2, BS3 and BS4 are beam splitters with transmissivities ${\rm T_1}$, ${\rm T_2}$, ${\rm t_{2n-1}}$, and ${\rm t_{2n}}$, respectively. PD1, PD2, PD3 and PD4: photo detectors. The operation is successfully achieved under the detection of a single photon at only one of two detectors PD1 and PD2 and the detection of a single-photon at only one of two detectors PD3 and PD4.}
\label{FIG_4}
\end{figure}

{\bf (iii)} In Sec. 3, we have seen that the optimal PNES $\sum^N_{n=0}C_n|n\rangle_a|n\rangle_b$ with $N=2$ for CV teleportation has the coefficients
$C_0\approx 0.765$, $C_1\approx 0.535$, and $C_2\approx 0.359$.
Under our first scheme, these coefficients can be obtained using the experimental parameters, e.g.
$s_1=s_2=0.1$, $\phi_1=0$, $\phi_2=\pi$, $r_1\approx 0.4589$, and $r_2\approx 0.9984$, with
$t_1=(1-r^2_1)^{1/2}$ and $t_2=-(1-r^2_2)^{1/2}$.
On the other hand, under the second scheme, the same coefficients can be obtained using the parameters ${\rm r_2\approx 0.3}$, ${\rm r_3\approx 0.3863}$ and ${\rm r_4\approx 0.6193}$, with ${\rm t_2=-(1-r^2_2)^{1/2}}$, ${\rm t_3=(1-r^2_3)^{1/2}}$, and ${\rm t_4=(1-r^2_4)^{1/2}}$.
Note that in our first scheme, we generate such a PNES using the NDPAs in the weak squeezing regime, $s=0.1$ ($0.869~dB$).
Furthermore, in the second scheme, we can obtain the same PNES only by adjusting the beam-splitter parameters, therefore, a high-level of squeezing is not necessary in our schemes. This is also true for the case of nonlocality test shown in Sec. 3. Under the first scheme, the optimized coefficients of the PNES for the nonlocality test are obtained using the parameters
$s_1=s_2=0.1$, $\phi_1=0$, $\phi_2=\pi$, $r_1\approx 0.38$ and $r_2\approx 0.999$, with
$t_1=(1-r^2_1)^{1/2}$ and $t_2=(1-r^2_2)^{1/2}$. Under the second scheme, they are obtained
using the parameters ${\rm r_2\approx 0.3}$, ${\rm r_3\approx 0.391}$ and ${\rm r_4\approx 0.221}$, with
${\rm t_2=(1-r^2_2)^{1/2}}$, ${\rm t_3=(1-r^2_3)^{1/2}}$, and ${\rm t_4=(1-r^2_4)^{1/2}}$.

\section{Experimental feasibility}\label{Experimental feasibility}
In order to further illustrate the feasibility of our proposed schemes, we address realistic experimental conditions in producing a PNES up to one-photon correlation, $C_0|0\rangle_a|0\rangle_b+C_1|1\rangle_a|1\rangle_b$, or two-photon correlation, $C_0|0\rangle_a|0\rangle_b+C_1|1\rangle_a|1\rangle_b+C_2|2\rangle_a|2\rangle_b$, as examples. In the two schemes of Figs. 3 and 4, we particularly consider each photodetector as an on-off detector that only distingushes two events, detection and non-detection, with efficiency $\eta$. The photodetection can then be characterized by a positive operator-valued measure (POVM) \cite{36-1, 36-2, 37} with two components $\hat{\Pi}_0=\sum_n (1-\eta)^n|n\rangle\langle n|$ (no click) and $\hat{\Pi}_1=\hat{I}-\hat{\Pi}_0$ (click).

In the first scheme, an arbitrary input state goes through a sequence of operations---a two-mode squeezing, a second-order superposition $t\hat{a}\hat{a}^{\dag}+r\hat{a}^{\dag}\hat{a}$ heralded by nonideal on-off detectors, and the inverse two-mode squeezing.
This sequence yields an output state
\begin{equation}
\rho_{\rm out}=\frac{{\rm Tr}_{cde}[\hat{\Pi}^c_0\hat{\Pi}^d_1\hat{\Pi}^e_1\hat{U}_1\rho_{\rm in}\hat{U}^{\dag}_1]}{{\rm Tr}_{abcde}[\hat{\Pi}^c_0\hat{\Pi}^d_1\hat{\Pi}^e_1\hat{U}_1\rho_{\rm in}\hat{U}^{\dag}_1]},
\end{equation}
where $\rho_{\rm in}\equiv |\psi\rangle\langle\psi |_{ab}\otimes |0\rangle\langle 0|_{cde}$ and
$\hat{U}_1\equiv \hat{S}^{\dag}_{ab}\hat{B}_{cd}\hat{B}_{ad}\hat{S}_{ae}\hat{B}_{ac}\hat{S}_{ab}$.
We can evaluate the performance of our scheme by investigating the fidelity $F_N=\langle\psi_N|\rho^{(N)}_{\rm out}|\psi_N\rangle$ between the ideal target state, $|\psi_N\rangle=\sum_{n=0}^NC_n|n\rangle_a|n\rangle_b$ ($N=1,2$)  and the corresponding output state, $\rho^{(N)}_{\rm out}$ ($N=1,2$).
In Fig. 5, we first show the case of PNES up to one-photon correlation, where the fidelity $F_1$ (blue dot) is plotted as a function of $|C_0|^2$ with the detector efficiency $\eta=0.66$.
With the parameters $s_1=0.1$ and $T^2_1=T^2_2=0.99$ in Fig. 3, we see that a high fidelity above $0.996$ is achieved in the whole range of $|C_0|^2$, with the detector efficiency $\eta=0.66$ currently available \cite{38,39,40}. 
For comparison, we plot the fidelity of the output state using the generalied scissor scheme of \cite{35-1} with the input two-mode squeezed state ($s=0.1$) and the on-off detectors ($\eta=0.66$). We see that our schemes yield a slightly better fidelity than the scissor scheme.

In Fig. 6, we further investigate the fidelity $F_2$ for the case of PNES up to two-photon correlation as a function of $|C_1|^2$ and $|C_2|^2$. With the same parameters ($\eta=0.66$, $s_1=0.1$ and $T^2_1=T^2_2=0.99$) used in Fig. 5, the fidelity is achieved at least above $0.941$ in the whole range of $|C_1|^{2}$ and $|C_2|^{2}$. For both of the cases, the fidelity slightly increases with the vacuum-state probability $|C_0|^2$, as the weak coupling ($s_1=0.1$) of the NDPA makes a low photon-number state better controlled.

\begin{figure}
\centering\includegraphics[scale=0.8]{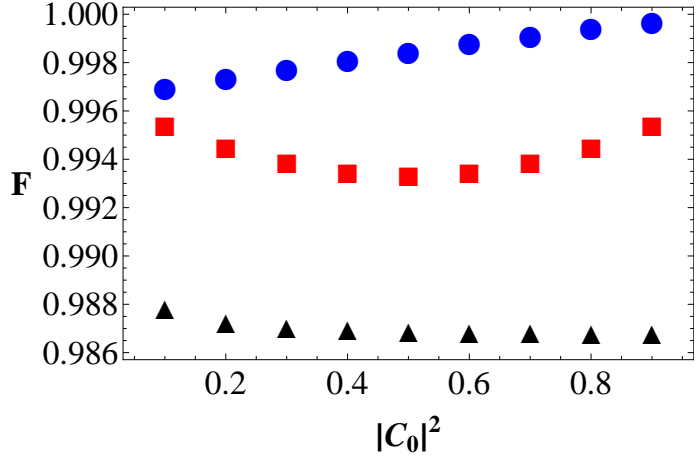}
\caption{Fidelity between the ideal state $C_0|0\rangle_a|0\rangle_b+C_1|1\rangle_a|1\rangle_b$ and the output state $\rho_{\rm out}$
obtained by applying $\hat{S}^{\dag}_{ab}(\xi)(t\hat{a}\hat{a}^{\dag}+r\hat{a}^{\dag}\hat{a})\hat{S}_{ab}(\xi)$ (blue circle) or
$({\rm t_2}\hat{a}+{\rm r_2}\hat{b}^{\dag})({\rm t_1}\hat{b}+{\rm r_1}\hat{a}^{\dag})$ (red square), using
on-off detectors with efficiency $\eta$ to the input state $\rho_{\rm in}=|0\rangle_a|0\rangle_b$ as a function of $|C_0|^2$ for $\eta=0.66$. Black triangle represents the output fidelity using the scissor scheme of \cite{35-1}, with the input two-mode squeezed state ($s=0.1$) and the on-off detectors ($\eta=0.66$).}
\label{FIG_6}
\end{figure}

In the second scheme, two first-order superposition operations $({\rm t_2}\hat{a}+{\rm r_2}\hat{b}^{\dag})({\rm t_1}\hat{b}+{\rm r_1}\hat{a}^{\dag})$ heralded by nonideal on-off detectors are sequentially applied to an arbitrary input state.
This yields an output state
\begin{equation}
\rho^{'}_{\rm out}=\frac{{\rm Tr}_{cdef}[\hat{\Pi}^e_0\hat{\Pi}^f_1\hat{\Pi}^c_0\hat{\Pi}^d_1\hat{U}_2\rho^{'}_{\rm in}\hat{U}^{\dag}_2]}{{\rm Tr}_{abcdef}[\hat{\Pi}^e_0\hat{\Pi}^f_1\hat{\Pi}^c_0\hat{\Pi}^d_1\hat{U}_2\rho^{'}_{\rm in}\hat{U}^{\dag}_2]},
\end{equation}
where $\rho^{'}_{\rm in}\equiv |\psi\rangle\langle\psi |_{ab}\otimes |0\rangle\langle 0|_{cdef}$ and
$\hat{U}_2\equiv \hat{B}_{ef}\hat{B}_{ae}\hat{S}_{bf}\hat{B}_{cd}\hat{B}_{bd}\hat{S}_{ac}$.
We investigate the fidelity $F^{'}_{N}=\langle\psi^{'}_{N}|\rho^{'(N)}_{\rm out}|\psi^{'}_{N}\rangle$ between the ideal target state,$|\psi_N'\rangle=\sum_{n=0}^NC_n|n\rangle_a|n\rangle_b$ ($N=1,2$),  and the corresponding output state, $\rho^{'(N)}_{\rm out}$ ($N=1,2$). 
In Fig. 5, we plot the fidelity $F^{'}_{1}$ (red square) as a function of $|C_0|^2$ with the detector efficiency $\eta=0.66$. With the parameters $\rm s_1=s_2=0.1$ and $\rm T^2_1=T^2_2=0.99$ in Fig. 4, we find that a high fidelity above $0.993$ is achieved in the whole range of $|C_0|^2$.
In Fig. 6, we investigate the fidelity $F^{'}_{2}$ as a function of $|C_1|^2$ and $|C_2|^2$. With the same parameters ($\eta=0.66, \rm s_1=s_2=0.1$ and $\rm T^2_1=T^2_2=0.99$) used in Fig. 5, the fidelity is achieved at least above $0.949$ in the whole range of $|C_1|^{2}$ and $|C_2|^{2}$. 
Therefore, both of our schemes seem to make an output state at a very high fidelity even with nonideal on-off detectors used for heralding the conditional generation of the PNES.

\begin{figure}
\centering\includegraphics[scale=0.8]{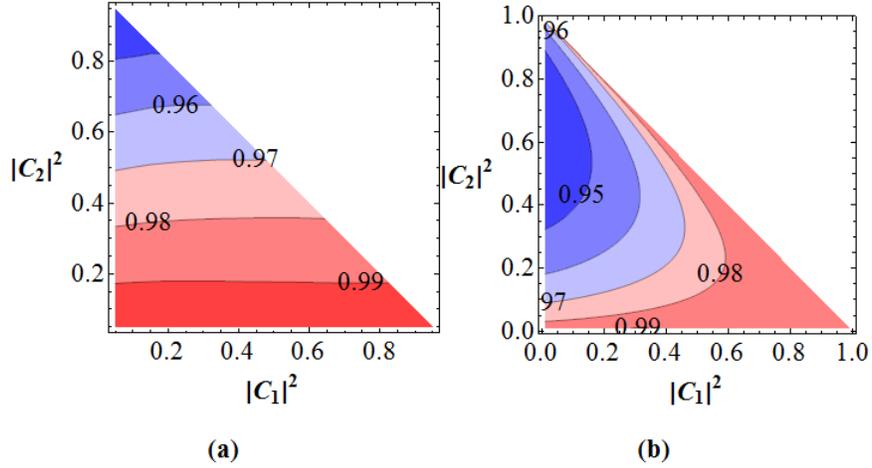}
\caption{Fidelity between the ideal state $C_0|0\rangle_a|0\rangle_b+C_1|1\rangle_a|1\rangle_b+C_2|2\rangle_a|2\rangle_b$ and the output state $\rho_{\rm out}$
obtained by applying twice (a) $\hat{S}^{\dag}_{ab}(\xi_2)(t_{2}\hat{a}\hat{a}^{\dag}+r_{2}\hat{a}^{\dag}\hat{a})\hat{S}_{ab}(\xi_2)\hat{S}^{\dag}_{ab}(\xi_1)(t_{1}\hat{a}\hat{a}^{\dag}+r_{1}\hat{a}^{\dag}\hat{a})\hat{S}_{ab}(\xi_1)$ or (b) $({\rm t_4}\hat{a}+{\rm r_4}\hat{b}^{\dag})({\rm t_3}\hat{b}+{\rm r_3}\hat{a}^{\dag})({\rm t_2}\hat{a}+{\rm r_2}\hat{b}^{\dag})({\rm t_1}\hat{b}+{\rm r_1}\hat{a}^{\dag})$, using
on-off detectors with efficiency $\eta$ to the input state $\rho_{\rm in}=|0\rangle_a|0\rangle_b$ as a function of $|C_1|^2$ and $|C_2|^2$ for $\eta=0.66$.}
\end{figure}

We have also calculated the success probability numerically for the output states under each scheme. For the state $C_0|0\rangle_a|0\rangle_b+C_1|1\rangle_a|1\rangle_b$, the first scheme, with the condition of $s_1=0.1$, $s=0.1$ and $T^2_1=T^2_2=0.99$, yields the success rate in the range of $2.4\times10^{-6}$ ($|C_0|^2=1/2$) to $10^{-4}$, which increases with the coefficient $|C_0|$. On the other hand, the second scheme, with the condition of $\rm s_1=s_2=0.1$ and $\rm T^2_1=T^2_2=0.99$, yields the success rate $\sim10^{-4}$.  The success probability can of course be made larger by using higher-squeezing NDPAs in each scheme at the expense of output fidelity to some extent. 

Other than nonideal detector efficiency, dark counts might potentially degrade the output fidelity. However, a recent experiment reported that a coincidence detection scheme recording only the synchronized events of laser pulse and a detector click in the pulsed regime can significantly eliminate dark count events \cite{41}. Another experimental imperfection may also arise from the error in the transmissivity of beam splitter under our proposed schemes. In Fig. 7, we plot the output fidelity from each scheme by including the error $\Delta t$ of beam-splitter transmissivity. Compared with Fig. 6, it turns out that a high output fidelity is still achievable and that the second scheme is particularly insensitive to the beam-splitter error.

\begin{figure}
\centering\includegraphics[scale=0.8]{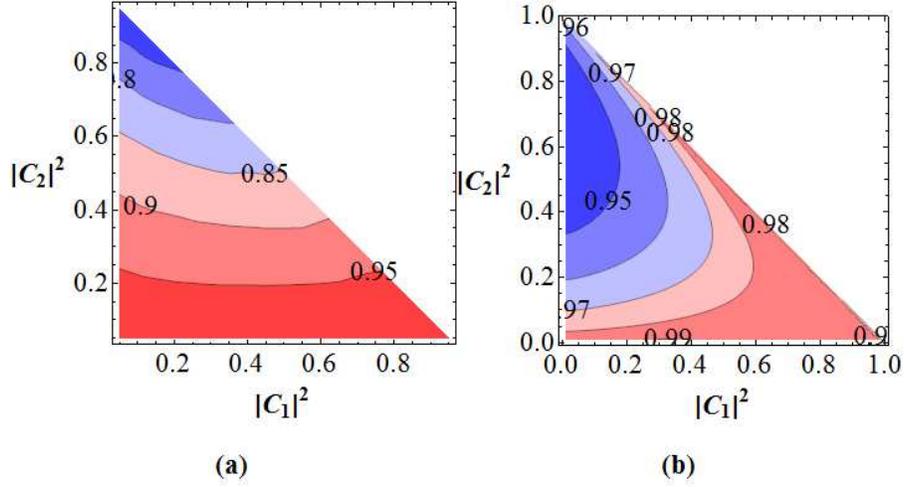}
\caption{Fidelity between the ideal state 
$C_0|0\rangle_a|0\rangle_b+C_1|1\rangle_a|1\rangle_b$ and the output state with the error $\Delta t_i =\pm 0.01$ of the beam-splitter transmissivity ($i=1,2$). 
Other parameters are the same as those in Fig. 6.}
\end{figure}

\section{Summary}\label{Summary and Remarks}
We have proposed two experimental schemes to generate a finite-dimensional photon number entangled state (PNES) with arbitrary coefficients, i.e., $\sum^N_{n=0}C_n|n\rangle_a|n\rangle_b$. One scheme is based on the second-order coherent superposition operation with two-mode squeezing operations, and the other on two first-order
coherent superposition operations. We have shown that the coefficients of the PNES can be adjusted by the parameters of beam splitters and NDPAs in each scheme. In particular, our schemes do not require a high-level of squeezing for the nonlinear materials (NDPAs) and we further demonstrated that our schemes can generate the PNESs with high fidelity using realistic on-off photodetectors with nonideal efficiency. The class of PNES is useful for CV quantum informatics as we have considered its application to quantum teleportation and nonlocality test. We have shown that the PNES of finite dimension can surpass the performance of the TMSS with the level of squeezing currently available in the pulsed regime. Furthermore, the PNES includes a broad class of non-Gaussian entangled states together with the TMSS (a representative Gaussian entangled state), therefore, our schemes can also be used for fundamental tests of quantum physics, e.g. the robustness of Gaussian versus non-Gaussian entanglement under noisy environments \cite{13, 14-1, 14-2}.

\section*{Acknowledgments}\label{Acknowledgments}
S.Y.L. thanks S.W.Ji for a helpful discussion.
This work is supported by the NPRP grant 4-346-1-061 from Qatar National Research Fund.

\begin{thebibliography}{99}
\bibitem{31-1} A recent experiment achieved a higher-squeezing level $\sim$6.8dB of a pulsed light at the wavelength $\lambda=$1500nm in optical fiber \cite{31-2}. For a long-distance quantum communication, however, one may require a quantum memory to store the quantum state of light. For this purpose, alkali atoms have been employed with the wavelength range $\lambda\sim$800nm, e.g. \cite{31-3}.
Furthermore, the thermal photon noise that can be detrimental to the quantum nature of light usually increases with the wavelength, so we here compare the PNES with the pulsed squeezed light of $\lambda=$850nm reported in \cite{31}.
\end{thebibliography}
\end{document}